\documentstyle[12pt,epsf]{ioplppt} 

\eqnobysec

\begin{document} 
\title{Complete phase diagram for the integrable chain with alternating 
spins in the sectors with competing interactions}[Phase diagram for 
XXZ($\frac{1}{2},1$)] \author{B - D D\"orfel\footnote{E-mail:
doerfel@qft2.physik.hu-berlin.de} and St Mei\ss ner\footnote{E-mail:
meissner@qft2.physik.hu-berlin.de}} \address{Institut f\"ur Physik,
Humboldt-Universit\"at , Theorie der Elementarteilchen\\
Invalidenstra\ss e 110, 10115 Berlin, Germany}

\begin{abstract} 
We investigate the anisotropic integrable spin chain consisting of
spins $s=\frac{1}{2}$ and $s=1$ by means of thermodynamic Bethe ansatz 
for the anisotropy $\gamma>\pi/3$, where the analysis of the 
Takahashi conditions leads to a more complicated string picture. We give
the phase diagram with respect to the two real coupling constants $\bar{c}$ and 
$\tilde{c}$, which contains a new region where the ground state is formed by 
strings with infinite Fermi zones. In this region the velocities of sound for 
the two physical excitations have been calculated from the dressed energies. 
This leads to an additional line of conformal invariance not known before. 

\end{abstract} 

\pacs{75.10 JM, 75.40 Fa}

\maketitle 
\section{Introduction} 
In 1992 de Vega and Woynarovich constructed the first example of an integrable
spin chain with alternating spins of the values $s=\frac{1}{2}$ and $s=1$ 
\cite{devega} on the basis of the well-known $XXZ(\frac{1}{2})$ model. We call 
this model $XXZ(\frac{1}{2},1)$. It contains two real coupling constants 
$\bar{c}$ and $\tilde{c}$. Most authors have limited their study to positive 
values for them. In our series of papers \cite{meissner,doerfel1,doerfel2} we
have studied the $XXZ(\frac{1}{2},1)$ model in the whole 
$(\bar{c},\tilde{c})$-plane and determined the ground state by thermodynamic 
Bethe ansatz (TBA) for equal signs of couplings and for competing interactions in 
the case $\gamma\leq\pi/3$. We have found four regions with different ground states. 
Two of them contain only strings with infinite Fermi zones, they include the sectors
with equal signs of the couplings and are well studied.

In this paper we want to deal with the remaining case of competing interactions
for $\gamma>\pi/3$. The paper is organized as follows. After having reviewed the 
definitions in section 2 we start our analysis with TBA in section 3. We restrict 
ourselves to special values of the anisotropy $\gamma>\pi/3$ and perform the analysis 
of the Takahashi conditions. The TBA equations are given explicitly. In section 4 we 
discuss them with respect to the values of the couplings $\bar{c}$ and $\tilde{c}$, 
what leads to the ground state phase diagram, which contains a new region with strings
having infinite Fermi zones. In this region the velocities of sound for the two 
physical excitations are calculated in section 5, while section 6 contains our 
conclusions.

We found it necessary to use the abbreviations I, II and III for our papers 
\cite{meissner}, \cite{doerfel1}, \cite{doerfel2} respectively.

\section{Description of the model}
We refer the reader to papers \cite{devega} and I for the basics 
of the model.

Our Hamiltonian of a spin chain of length $2N$ is given by
\begin{equation}\label{ham}
{\cal H}(\gamma) = \bar{c} \bar{\cal H}(\gamma) + \tilde{c} \tilde{\cal H}
(\gamma),
\end{equation}
with the two couplings $\tilde{c}$ and $\bar{c}$. The anisotropy parameter 
$\gamma$ is limited to $0<\gamma<\pi/2$. For convenience we repeat the Bethe 
ansatz equations (BAE), the magnon energies and momenta and the spin projection.
\begin{equation}\label{bae}
\fl \left( \frac{\sinh(\lambda_j+i\frac{\gamma}{2})}{\sinh(\lambda_j-i
\frac{\gamma}{2})}
\frac{\sinh(\lambda_j+i\gamma)}{\sinh(\lambda_j-i\gamma)} \right)^N =
-\prod_{k=1}^{M}\frac{\sinh(\lambda_j-\lambda_k+i\gamma)}{\sinh(\lambda_j-
\lambda_k-i\gamma)},\qquad j=1\dots M,
\end{equation}
\begin{eqnarray}\label{en}
E = \bar{c} \bar{E} + \tilde{c} \tilde{E},
\nonumber\\
\bar{E} = - \sum_{j=1}^{M} \frac{2\sin\gamma}
{\cosh2\lambda_j - \cos\gamma},
\nonumber\\
\tilde{E} = - \sum_{j=1}^{M} 
\frac{2\sin2\gamma}{\cosh2\lambda_j - \cos2\gamma},
\end{eqnarray}
\begin{equation}\label{mom}
P =\frac{i}{2}\sum_{j=1}^{M} \left\{ \log \left(\frac{\sinh(\lambda_j+i\frac{
\gamma}{2})}{\sinh(\lambda_j-i\frac{\gamma}{2})} \right) + 
\log \left( \frac{\sinh(\lambda_j+i\gamma)}{\sinh(\lambda_j-i\gamma)} \right) 
\right\},
\end{equation}
\begin{equation}\label{spin}
S_z = \frac{3N}{2} - M.
\end{equation}

\section{Thermodynamic Bethe ansatz (TBA)}
Following our paper I we assume that the solutions of (\ref{bae}) 
are of the string-type in the thermodynamic limit
\begin{equation}\label{string}
\lambda_{\alpha}^{n,j,\nu} = \lambda_{\alpha}^{n,\nu} + i(n+1-2j)\frac{\gamma}
{2} + \frac{1}{4} i \pi (1-\nu) + \delta_{\alpha}^{n,j,\nu},\qquad j=1\dots n.
\end{equation}
Here $\lambda_{\alpha}^{n,\nu}$ is the real center of the string, $n$ is the
string length and $\nu$ the parity of the string with values $\pm1$. The last
term is a correction due to finite size effects. These strings have to obey the 
Takahashi conditions \cite{taka}
\begin{equation}\label{cond}
\nu_n \sin \gamma j \sin \gamma (n-j) > 0, \qquad j=1 \dots n-1.
\end{equation}
Substituting (\ref{string}) into (\ref{bae}) and taking the logarithm yields
\begin{eqnarray}\label{logbae}
N t_{j,1}(\lambda_{\alpha}^{n_j}) + N t_{j,2}(\lambda_{\alpha}^{n_j}) 
= 2\pi I_{\alpha}^{n_j} + \sum_{k} \sum_{\beta} \Theta_{jk}(\lambda_{\alpha}
^{n_j}-\lambda_{\beta}^{n_k},\nu_j\nu_k)
\end{eqnarray}
with the known notations
\begin{equation}
t_{j,2S}(\lambda) = \sum_{k=1}^{\mbox{min}(n_j,2S)} 
f(\lambda,|n_j-2S|+2k-1,\nu_j),
\end{equation}
\begin{eqnarray}
\fl
\Theta_{jk}(\lambda) = f(\lambda,|n_j-n_k|,\nu_j\nu_k) 
\nonumber\\
+ f(\lambda,(n_j+n_k),\nu_j\nu_k) 
+ 2 \!\!\!\!\!\!\!\sum_{k=1}^{min(n_j,n_k)-1} 
\!\!\!\!f(\lambda,|n_j-n_k|+2k,\nu_j\nu_k),
\end{eqnarray}
and
\begin{eqnarray}
f(\lambda,n,\nu) = \left\{ \begin{array}{r@{\quad\quad}l}
					0 & n\gamma/\pi \in \bold Z\\
					2\nu \arctan ((\cot (n\gamma/2))^{\nu}
				\tanh\lambda) & n\gamma/\pi \notin \bold Z
			  \end{array} \right. .
\end{eqnarray}
Here we have used that a given string length $n>1$ corresponds to a unique 
parity, what is a consequence of (\ref{cond}). The numbers $I_{\alpha}^{n_j}$ 
are half-odd-integers counting the strings of length $n_j$.

Introducing particle and hole densities in the usual way we perform the limiting
process $N\to\infty$
\begin{eqnarray}\label{intbae}
a_{j,1}(\lambda) + a_{j,2}(\lambda)
= (\rho_j(\lambda)+\tilde{\rho}_j(\lambda))(-1)^{r(j)}+
\sum_{k} T_{jk} * \rho_k(\lambda),
\end{eqnarray}
where $a*b(\lambda)$ denotes the convolution
\begin{equation}
a*b(\lambda) = \int_{-\infty}^{\infty}d\mu a(\lambda-\mu)b(\mu)
\end{equation}
and
\begin{equation}
a_{j,2S}(\lambda) = \frac{1}{2\pi}\frac{d}{d\lambda}t_{j,2S}(\lambda),
\qquad
T_{j,k}(\lambda) =  \frac{1}{2\pi}\frac{d}{d\lambda}\Theta_{j,k}(\lambda).
\end{equation}

The sign $(-1)^{r(j)}$ results from the requirement of positive densities in 
the 'non-interacting' limit.

We are now able to express energy, momentum and spin in terms of the densities 
via (\ref{en}), (\ref{mom}) and (\ref{spin}). The standard procedure leads to 
equations determining the equilibrium state at temperature $T$ (TBA equations):
\begin{eqnarray}\label{tbae}
\fl
T\ln\left(1+\exp\left(\frac{\epsilon_j}{T}\right)\right) = 
\nonumber\\
-2\pi\bar{c}a_{j,1}(\lambda) - 2\pi\tilde{c} a_{j,2}(\lambda)
+ \sum_{k} T\ln\left(1+\exp\left(\frac{-\epsilon_k}{T}\right)\right) 
* A_{jk}(\lambda)
\end{eqnarray}
with
\begin{equation}
A_{jk}(\lambda)=(-1)^{r(k)}T_{jk}(\lambda,\nu_j\nu_k) + \delta(\lambda)
\delta_{jk}
\end{equation}
and
\begin{equation}
\frac{\tilde{\rho}_j}{\rho_j}=\exp\left(\frac{\epsilon_j}{T}\right).
\end{equation}
Again the free energy can be expressed in terms of our new variables 
$\epsilon_j(\lambda)$:
\begin{equation}\label{free}
\fl
2 {\cal F} = \frac{F}{N} = -\int_{-\infty}^{\infty} d\lambda \sum_{j}
(-1)^{r(j)}(a_{j,1}(\lambda) + a_{j,2}(\lambda))T\ln\left(1+\exp\left(\frac
{-\epsilon_j}{T}\right)\right).
\end{equation}

In I we analysed TBA equations for $\gamma=\pi/\mu,\quad\mu\dots$ 
integer, $\mu\geq 3$, where strings
\begin{enumerate}
\item
$n_j=j,\qquad\nu_j=1,\qquad j=1\dots\mu-1$,
\item
$n_{\mu}=1,\qquad\nu_{\mu}=-1$
\end{enumerate}
occur. The equations obtained allow the complete discussion of the ground state 
properties for values $0<\gamma\leq\pi/3$. On the other hand the picture for 
$\gamma>\pi/3$ is still not fully clear. For this reason we want to investigate
the case
\begin{equation}
\frac{\pi}{\gamma}=2+\frac{1}{\mu},\quad\mu\in{\bold N},\quad\mu\geq2.
\end{equation}
This restricts $\gamma$ to $2\pi/5<\gamma<\pi/2$. The analysis of (\ref{cond}) then 
leads to three Takahashi zones with
\begin{enumerate}
\item
$n_1=1,\qquad\nu_1=1$,
\item
$n_j=2j-3,\qquad\nu_j=(-1)^{j+1},\qquad j=2\dots\mu+1$,
\item
$n_{\mu+2}=2,\qquad\nu_{\mu+2}=1$.
\end{enumerate}

The operator $A_{jk}$ (\ref{tbae}) now has to be reversed by 
\begin{eqnarray}
C_{11}=\delta(\lambda)-d_1(\lambda),\nonumber\\
C_{21}=-C_{23}=s_2(\lambda),\quad C_{22}=1,\nonumber\\
C_{jk} = \delta(\lambda)\delta_{jk} - s_2(\lambda)(\delta_{j+1k}+\delta_{j-1k}),
\qquad j,k=3\dots\mu,\nonumber\\
C_{\mu+1,\mu}=-s_2(\lambda),\quad 
C_{\mu+1\mu+1}=C_{\mu+2\mu+1}=-C_{\mu+1\mu+2}=\frac{1}{2}\delta(\lambda),
\nonumber\\
C_{\mu+2\mu+2}=\delta(\lambda),
\end{eqnarray}
where
\begin{eqnarray}
\fl
s_2(\lambda) = \frac{1}{2\gamma\cosh(\pi\lambda/(\pi-2\gamma))},\quad
d_1(\lambda)=\frac{1}{2\pi}\int_{-\infty}^{\infty}e^{i\omega\lambda}
\frac{\cosh(\omega(\pi-3\gamma)/2)}{2\cosh(\omega(\pi-2\gamma)/2)
\cosh(\omega\gamma/2)}.
\nonumber\\
\end{eqnarray}
The reversed TBA equations read
\begin{eqnarray}\label{invtbae2}
\fl
\epsilon_1(\lambda) &=& - 2\pi\bar{c} s_1(\lambda)- 2\pi\tilde{c} d_1(\lambda)
- T d_1*\ln(f(\epsilon_1)) - T s_2*\ln(f(\epsilon_2)),
\nonumber\\
\fl
\epsilon_2(\lambda) &=& 2\pi\tilde{c} s_2(\lambda) 
+ T s_2*\ln(f(\epsilon_3)) - T s_2*\ln(f(\epsilon_1)) ,
\nonumber\\
\fl
\epsilon_j(\lambda) &=& - Ts_2*\ln(f(\epsilon_{j+1})f(\epsilon_{j-1})) 
- \delta_{j\mu}
Ts_2*\ln(f(-\epsilon_{\mu+2})),\quad j=3\dots\mu,
\nonumber\\
\fl
\epsilon_{\mu+1}(\lambda) &=& -Ts_2*\ln(f(\epsilon_{\mu})),
\nonumber\\
\fl
\epsilon_{\mu+2}(\lambda) &=& Ts_2*\ln(f(\epsilon_{\mu})).
\end{eqnarray}
with
\begin{equation}
s_1(\lambda) = \frac{1}{2\gamma\cosh(\pi\lambda/\gamma)}
\end{equation}
and  the Fermi function
\begin{eqnarray}
f(x)=\frac{1}{1+e^{x/T}}.
\end{eqnarray}

This system of equations looks very similar to equations (3.17) in I. In both 
cases only the strings with Takahashi indices $1$, $2$, $\mu+2$ ($\mu$ in I) can
occur in the ground state. Notice that the $(2,+)$-strings and the 
$(1,-)$-strings have interchanged their positions in I(3.17) and 
(\ref{invtbae2}).

\section{Ground states and phase diagram}
To obtain the ground state one has to carry out $T\to0$ in (\ref{tbae})
and (\ref{invtbae2}). Eliminating the strings which are not 
relevant for the ground state one arrives at the systems
\begin{eqnarray}\label{-eq}
\epsilon_{1+}(\lambda)=-2\pi\bar{c}s_1(\lambda) - K_1*\epsilon_{1-}^-(\lambda),
\nonumber\\
\epsilon_{2+}(\lambda)=-2\pi\tilde{c}s_1*s_1(\lambda) -2\pi\tilde{c}s_1(\lambda)
*\epsilon_{1+}^-(\lambda) + K_2*\epsilon_{1-}^-(\lambda),
\nonumber\\
\epsilon_{1-}(\lambda)= 2\pi\bar{c}s_1*K_1(\lambda) + 2\pi\tilde{c}K_1(\lambda)
+ K_1*\epsilon_{1+}^-(\lambda) -  K_3*\epsilon_{1-}^-(\lambda),
\end{eqnarray}
or equivalently
\begin{eqnarray}\label{+eq}
\epsilon_{1+}(\lambda) = - 2\pi\bar{c} s_1(\lambda)- 2\pi\tilde{c} d_1(\lambda)
+ d_1*\epsilon_{1+}^+(\lambda) + s_2*\epsilon_{1-}^+(\lambda),
\nonumber\\
\epsilon_{2+}(\lambda)= - K_4*\epsilon_{1-}^+(\lambda),
\nonumber\\
\epsilon_{1-}(\lambda)= 2\pi\tilde{c}s_2(\lambda) - s_2*\epsilon_{1+}^+(\lambda)
+(\delta-K_5)*\epsilon_{1-}^+(\lambda).
\end{eqnarray}
Here for the sake of clarity we have labelled the strings by their
lengths and parities instead of Takahashi indices. $\epsilon^{\pm}$ denote 
positive and negative parts of $\epsilon$ respectively. The functions newly 
introduced are defined via their Fourier transforms and are given in the 
appendix.

It is remarkable that these equations are valid for the whole region 
$\pi/3<\gamma<\pi/2$ though they are obtained relying on the TBA for 
$2\pi/5\leq\gamma<\pi/2$. This is due to the fact that the relevant part of the 
string picture does not change passing the point $\gamma=2\pi/5$, what can easily 
be checked by performing the Takahashi analysis for $1/3<\pi/\gamma<5/2$. 

\begin{figure}
\hbox{
\epsfxsize=190pt
\epsfbox{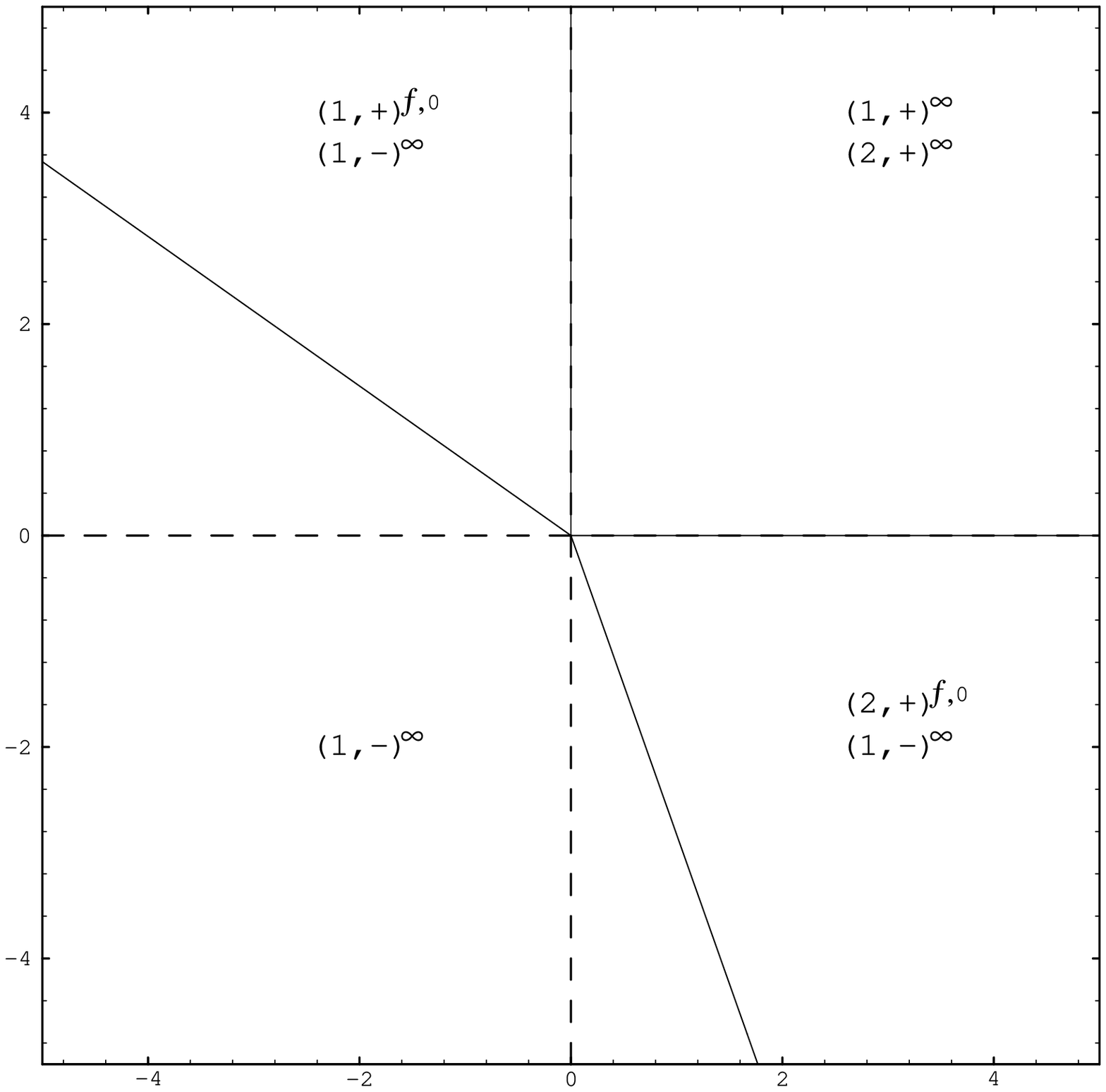}
\epsfxsize=190pt
\epsfbox{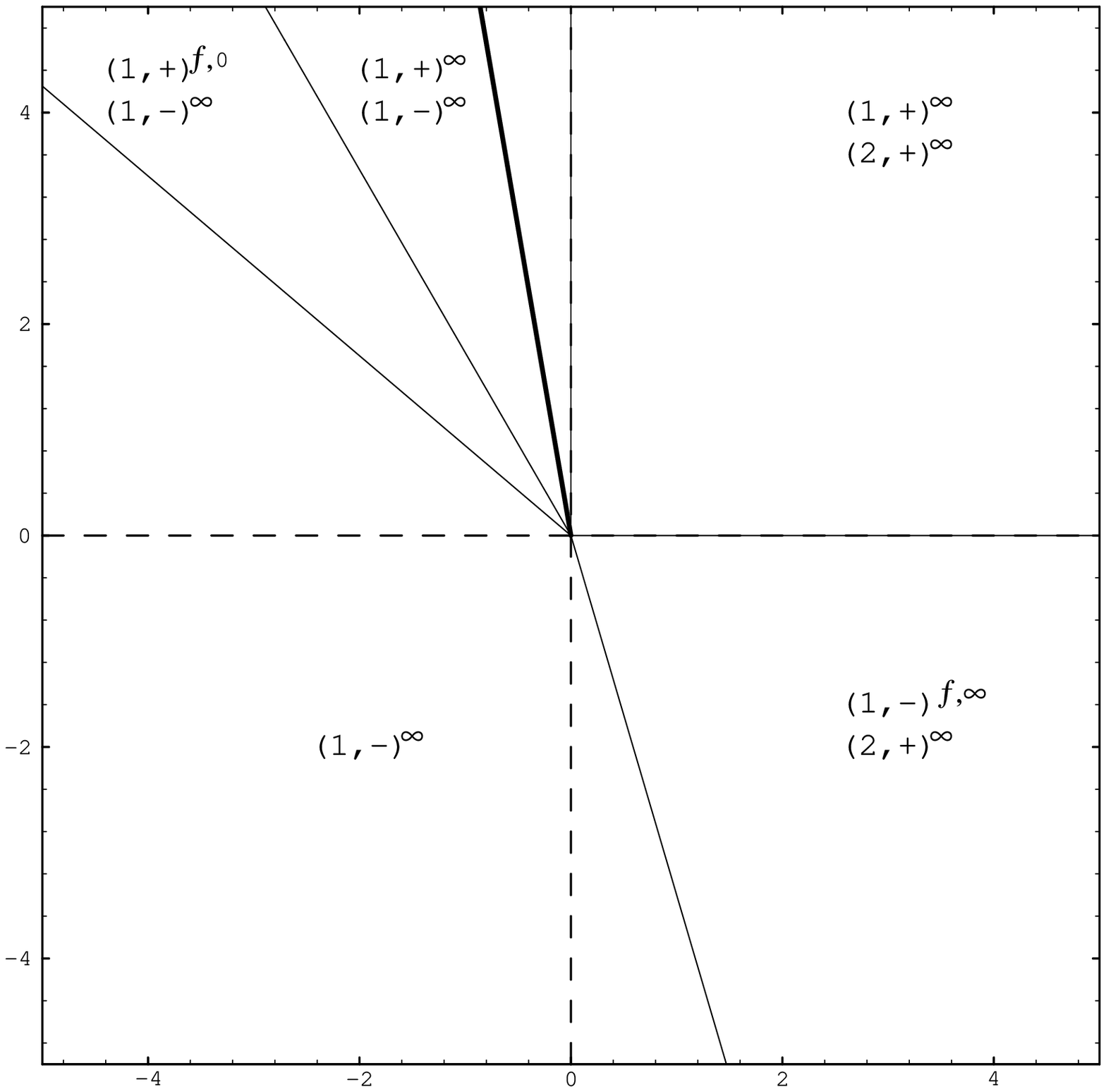}}
\hbox{
\epsfxsize=190pt
\epsfbox{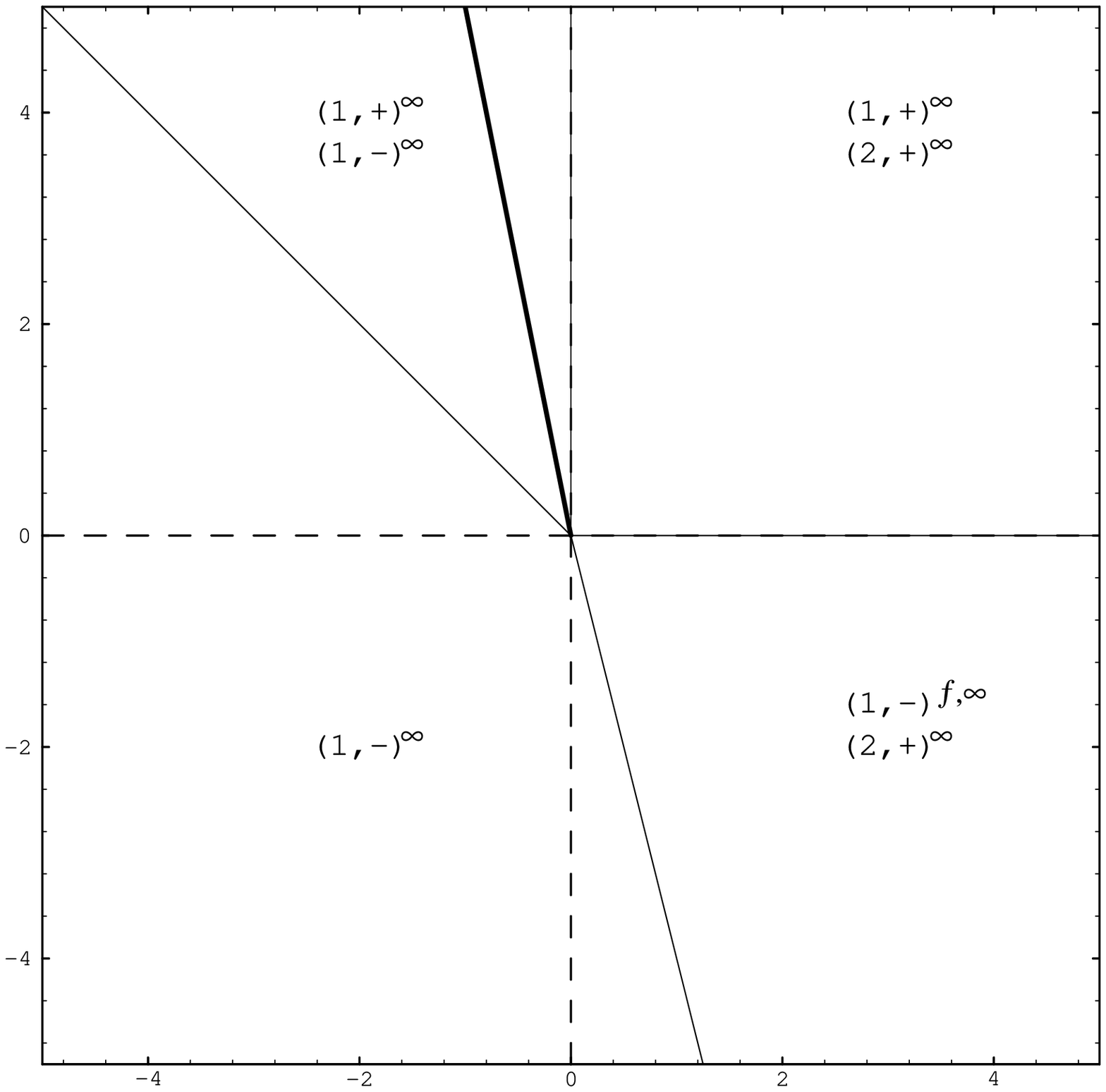}
\epsfxsize=190pt
\epsfbox{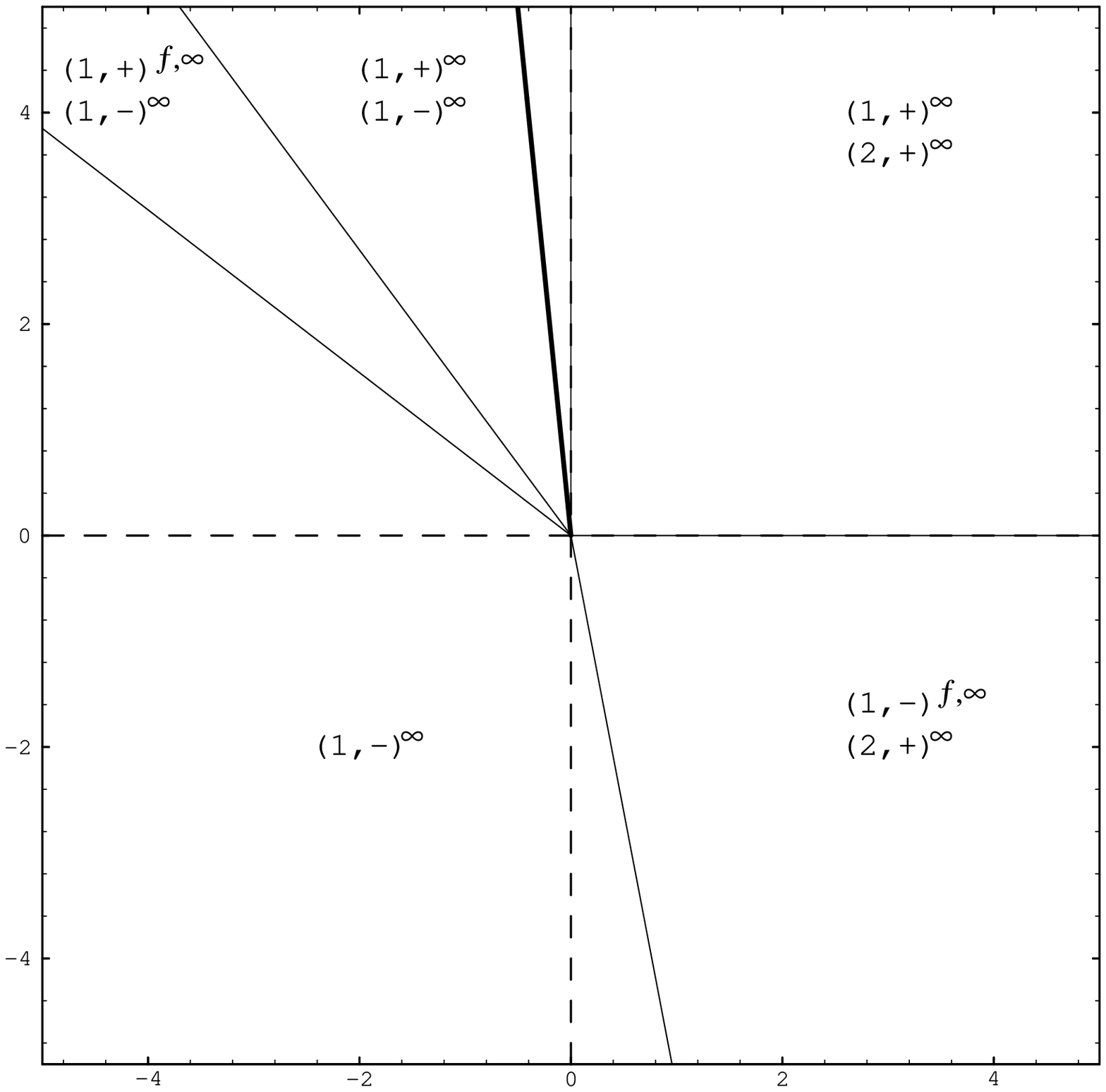}}
\caption{\label{pf}Phase diagram in the $(\tilde{c},\bar{c})$-plane for 
$\gamma=\pi/3$, $3\pi/8$, $2\pi/5$ and $3\pi/7$ respectively. The string contents of 
the sectors is indicated. Axes are drawn broken except they coincide with sector 
borders. Upper indices indicate infinite and finite Fermi zones. In the latter case, 
the second index distinguishes, wether the filling starts at $\lambda=0$ or 
$\lambda=\infty$. The bold line marks a new line of conformal invariance (see 
section 5).}
\end{figure}

Now one can discuss (\ref{-eq}) and (\ref{+eq}) with respect to the signs of
$\bar{c}$ and $\tilde{c}$ for $\pi/3<\gamma<\pi/2$.
\begin{itemize}
\item[a)]$\bar{c},\tilde{c}>0$

The solution can be given explicitly. We have
\begin{equation}
\epsilon_{1+}(\lambda) = -2\pi\bar{c}s_1(\lambda),\qquad
\epsilon_{2+}(\lambda) = -2\pi\tilde{c}s_1(\lambda).
\end{equation}
This is the solution of de Vega and Woynarovich \cite{devega}.

\item[b)]$\bar{c}<0$

From (\ref{-eq}) we have $\epsilon_{1+}\equiv\epsilon_{1+}^+$. Therefore we find
the following integral equations for the two relevant functions $\epsilon_{2+}$ 
and $\epsilon_{1-}$:
\begin{eqnarray}
\epsilon_{2+}(\lambda)= - K_4*\epsilon_{1-}^+(\lambda),
\nonumber\\
\epsilon_{1-}(\lambda)= 2\pi\bar{c}s(\lambda) + \tilde{c}g(\lambda,\gamma)
+ s*K_6*\epsilon_{1-}^+(\lambda).
\end{eqnarray}
(The definition of $g(\lambda,\gamma)$ is repeated in the appendix.)
From these equations it can be seen that for vanishing positive part of 
$\epsilon_{1-}(\lambda)$ the solution discussed in I is reproduced.
The borderlines of this sector are found in II. Beyond this line 
$\epsilon_{1-}(\lambda)$ is positive in an intervall $(-b,b)$. $b$ is called the 
Fermi radius. It increases moving counterclockwise towards the line $\bar{c}=0$,
where it reaches infinity (see figure \ref{pf}).  

\item[c)]$\tilde{c}<0$

From (\ref{+eq}) we see that the solution can only be selfconsistent, if 
$\epsilon_{1-}^+\equiv0$ holds. One is left with the integral equations
\begin{eqnarray}
\epsilon_{1+}(\lambda) = - 2\pi\bar{c} s_1(\lambda)- 2\pi\tilde{c} d_1(\lambda)
+ d_1*\epsilon_{1+}^+(\lambda),
\nonumber\\
\epsilon_{1-}(\lambda)= 2\pi\tilde{c}s_2(\lambda) -s_2*\epsilon_{1+}^+(\lambda)
\end{eqnarray}
for $\epsilon_{1+}$ and $\epsilon_{1-}$. We want to discuss the case 
$\epsilon_{1+}^+\equiv0$ first. Then the solution reads
\begin{eqnarray}\label{sol}
\epsilon_{1+}(\lambda) = - 2\pi\bar{c} s_1(\lambda)- 2\pi\tilde{c} d_1(\lambda),
\nonumber\\
\epsilon_{1-}(\lambda)= 2\pi\tilde{c}s_2(\lambda).
\end{eqnarray}
The sector where this solution exists is given by the requirements
\begin{eqnarray}
\epsilon_{1+}(\lambda)\leq0,\qquad
\epsilon_{1-}(\lambda)\leq0.
\end{eqnarray}
While the latter condition is always fulfilled we expect restrictions on the 
values of $\bar{c}$ and $\tilde{c}$ coming from the first one. Putting 
$\lambda=0$ yields
\begin{eqnarray}\label{in1}
\frac{\bar{c}}{|\tilde{c}|}\geq 2\gamma d_1(0).
\end{eqnarray}
For $\lambda\to\infty$ we look at the asymptotics
\begin{eqnarray}
\epsilon_{1+}(\lambda)\cong -\frac{2\pi}{\gamma} e^{-\pi\lambda/\gamma}
\left(\bar{c}+\tilde{c}\tan\left(\frac{\pi^2}{2\gamma}\right)\right).
\end{eqnarray}
This yields
\begin{eqnarray}\label{in2}
\frac{\bar{c}}{|\tilde{c}|}\geq\tan\left(\frac{\pi^2}{2\gamma}\right).
\end{eqnarray}
The more restrictive one of the inequalities (\ref{in1}) and (\ref{in2}) marks 
the borderline of the investigated sector. Noticing
\begin{equation}\label{re1}
\tan\left(\frac{\pi^2}{2\gamma}\right)>2\gamma d_1(0)\qquad\mbox{for}\quad
\frac{\pi}{3}<\gamma<\frac{2\pi}{5},
\end{equation}
\begin{equation}\label{re2}
\tan\left(\frac{\pi^2}{2\gamma}\right)<2\gamma d_1(0)\qquad\mbox{for}\quad
\frac{2\pi}{5}<\gamma<\frac{\pi}{2}
\end{equation}
and
\begin{equation}\label{re3}
2\gamma d_1(0)=1\qquad\mbox{for}\quad\gamma=\frac{2\pi}{5}
\end{equation}
we conclude the following: The region with infinite Fermi zones for 1-strings with
different parities is given by equation (\ref{in1}) for $2\pi/5\leq\gamma<\pi/2$
and by equation (\ref{in2}) for $\pi/3<\gamma\leq2\pi/5$. Below the line specified
by (\ref{in1}) and (\ref{in2}) respectively $\epsilon_{1+}$ is positive in an 
intervall $(-b,b)$ for $2\pi/5<\gamma<\pi/2$, while it is positive for 
$|\lambda|\in(b,\infty)$ for $\pi/3<\gamma<2\pi/5$. $b$ is again called the Fermi
radius. Moving counterclockwise in the $(\bar{c},\tilde{c})$-plane from the line
(\ref{in1}) or (\ref{in2}), $b$ increases (decreases) until $\epsilon_{1+}$ is
completely positive and the solution from I is reproduced (see figure \ref{pf}). 
The borderline of this sector is again given in II. For $\gamma=2\pi/5$ no region 
with finite Fermi zone for (1,+) exists. The phaselines coincide at this point 
(see figure \ref{pf}).
\end{itemize}

\begin{figure}
\epsfxsize=380pt
\epsfbox{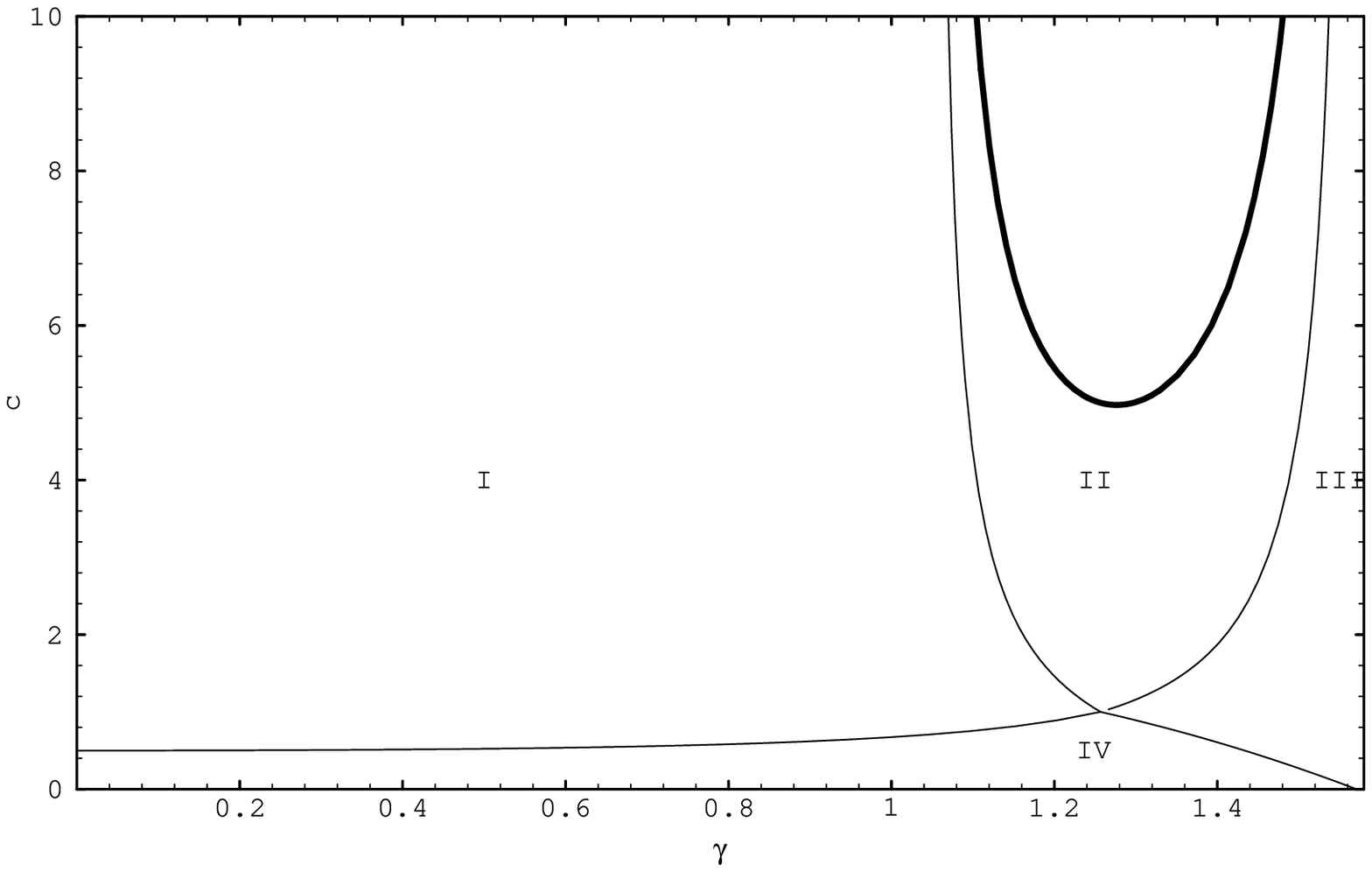}
\caption{\label{p+-}Phase diagram for $c=\bar{c}/|\tilde{c}|>0$ over $\gamma$. The 
sectors labelled differ with respect to their string-contents. The bold line 
indicates a new line of conformal invariance.}
\end{figure}

\begin{figure}
\epsfxsize=380pt
\epsfbox{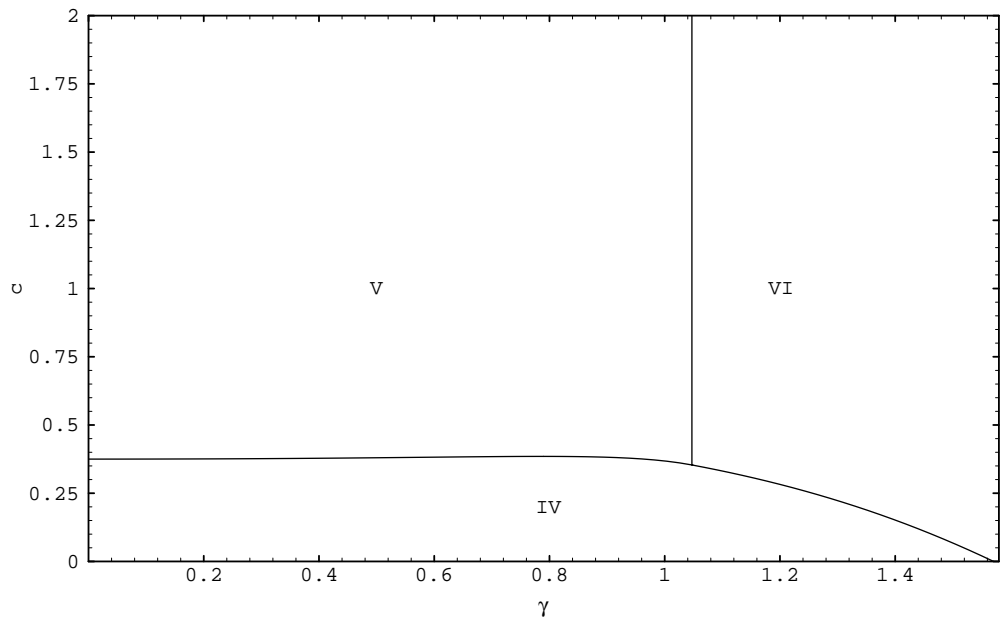}
\caption{\label{p2-}Phase diagram for $c=\tilde{c}/|\bar{c}|>0$ over $\gamma$.}
\end{figure}

This, together with the results from I and II, allows us to give the phase diagram 
in the sectors with competing interactions, i.e for different signs of the coupling
constants (see figures \ref{p+-} and \ref{p2-}). The sectors are labelled according 
to table \ref{t} and are different with respect to the string-contents. 

\begin{table}
\begin{center}
\begin{tabular}{|c||c|}\hline
I & $(1,-)^{\infty}$, $(1,+)^{f,0}$\\ \hline
II & $(1,-)^{\infty}$, $(1,+)^{\infty}$\\ \hline
III & $(1,-)^{\infty}$, $(1,+)^{f,\infty}$\\ \hline
IV  & $(1,-)^{\infty}$\\ \hline
V & $(1,-)^{\infty}$, $(2,+)^{f,0}$\\ \hline
VI & $(1,-)^{f,\infty}$, $(2,+)^{\infty}$\\ \hline
\end{tabular}
\end{center}
\caption{\label{t}Sectors appearing in the phase diagram for competing interactions. 
Upper indices indicate infinite and finite Fermi zones. In the latter case, 
the second index distinguishes, wether the filling starts at $\lambda=0$ or 
$\lambda=\infty$.}
\end{table}

\section{The new region II}
Next we deal with the region specified by equations (\ref{in1}) and (\ref{in2}). 
The dressed energies are given by (\ref{sol}). The dressed momenta can be easily 
found noticing
\begin{eqnarray}\label{dremom}
\frac{\d p(\lambda)}{\d \lambda} = - \left.\frac{\epsilon(\lambda)}{2}
\right|_{\bar{c}=\tilde{c}=1}.
\end{eqnarray}
Therefore from (\ref{sol}) the dispersion relation 
\begin{eqnarray}\label{disp-}
\epsilon_{1-}=\frac{2\pi\tilde{c}}{\pi-2\gamma}\frac{\sin 2 p_{1-}}{2}
\end{eqnarray}
follows with the speed of sound
\begin{eqnarray}\label{sos-}
v_{1-}=-\frac{2\pi\tilde{c}}{\pi-2\gamma}>0.
\end{eqnarray}
The dispersion relation for the $(1,+)$-strings is given implicitly by (\ref{sol})
together with (\ref{dremom}). The speed of sound then reads
\begin{eqnarray}\label{sos+}
v_{1+}=\frac{2\pi}{\gamma}
\frac{\bar{c}+\tilde{c}\tan(\pi^2/(2\gamma))}{1+\tan(\pi^2/(2\gamma))}\geq0.
\end{eqnarray}
It vanishes on the sector border (\ref{in2}) for $\pi/3<\gamma\leq2\pi/5$.

Now it is natural to look for possible lines of conformal invariance which must 
have $v_{1+}=v_{1-}$, with the solution
\begin{eqnarray}\label{cline}
\frac{\bar{c}}{|\tilde{c}|}=\frac{1}{\pi-2\gamma}\left[\gamma+(\pi-\gamma)
\tan \left(\frac{\pi^2}{2\gamma}\right)\right].
\end{eqnarray}
Analytical and numerical estimates show that there is always a solution fulfilling
(\ref{in1}) and (\ref{in2}).

\section{Conclusions}
We have investigated the $XXZ(\frac{1}{2},1)$ model in the region of anisotropy
$\pi/3<\gamma<\pi/2$ by means of TBA. The integral equations describing the ground 
state change drastically when passing the point $\gamma=\pi/3$. While in the case
of equal signs of the coupling constants this is of no influence on the ground state 
(what is already known from \cite{meissner},\cite{doerfel1}), in the case of competing
interactions the picture changes. The most striking consequence is the existence of 
a new region with strings $(1,+)$ and $(1,-)$ having infinite Fermi zones in the 
sector $\bar{c}>0,\tilde{c}<0$, which contains also a new line exhibiting conformal 
invariance. Outside this region, finite Fermi radius for the $(1,+)$-strings occurs. 
Here the model behaves similar to the one investigated in \cite{tsve1} and 
\cite{frahm1}.  

In the sector $\bar{c}<0,\tilde{c}>0$ the strings $(2,+)$ and $(1,-)$ interchange 
their behaviours compared to $\gamma\leq\pi/3$, i.e. now the $(1,-)$-strings occur with 
finite Fermi zone, while the $(2,+)$-strings always have an infinite one.

In sectors III and VI, where the filling for the $(1,+)$- respectively $(1,-)$-strings
starts at infinity, this causes the appearance of two different speeds of sound for each 
of them, to be calculated at $\lambda=b$ and $\lambda=\infty$.

It seems worthwhile to investigate these new regions in the sectors of competing
interaction with respect to the excitations. We hope to return to this point in
a future publication.

\section*{Appendix}
\begin{eqnarray*}
g(\omega,\alpha)=2\pi \frac{\cosh\omega\alpha/2}{\cosh\omega(\pi-\gamma)/2}\\
K_1(\omega)=\frac{\cosh\omega(\pi-3\gamma)/2}{\cosh\omega\gamma/2}\\
K_2(\omega)=\frac{\cosh\omega(\pi-\gamma)/2}{2\cosh^2\omega\gamma/2}\\
K_3(\omega)=\frac{\sinh^2\omega(\pi-2\gamma)/2}{\cosh^2\omega\gamma/2}\\
K_4(\omega)=\frac{1}{2\cosh\omega(\pi-3\gamma)/2}\\
K_5(\omega)=\frac{\cosh\omega\gamma/2}{2\cosh\omega(\pi-\gamma)/2
\quad\cosh\omega(\pi-3\gamma)/2}\\
K_6(\omega)=\frac{2\sinh^2\omega(\pi-2\gamma)/2}{\cosh\omega(\pi-3\gamma)/2}\\
s(\omega)=\frac{1}{2\cosh\omega(\pi-\gamma)/2}
\end{eqnarray*}

\newpage
\section*{Figure and table captions}
{\bf Figure 1.} Phase diagram in the $(\tilde{c},\bar{c})$-plane for 
$\gamma=\pi/3$, $3\pi/8$, $2\pi/5$ and $3\pi/7$ respectively. The string contents of 
the sectors is indicated. Axes are drawn broken except they coincide with sector 
borders. Upper indices indicate infinite and finite Fermi zones. In the latter case, 
the second index distinguishes, wether the filling starts at $\lambda=0$ or 
$\lambda=\infty$. The bold line marks a new line of conformal invariance (see 
section 5).\\[2cm]
{\bf Figure 2.} Phase diagram for $c=\bar{c}/|\tilde{c}|>0$ over $\gamma$. The 
sectors labelled differ with respect to their string-contents. The bold line 
indicates a new line of conformal invariance.\\[2cm]
{\bf Figure 3.} Phase diagram for $c=\tilde{c}/|\bar{c}|>0$ over $\gamma$.\\[2cm]
{\bf Table 1.} Sectors appearing in the phase diagram for competing interactions. 
Upper indices indicate infinite and finite Fermi zones. In the latter case, 
the second index distinguishes, wether the filling starts at $\lambda=0$ or 
$\lambda=\infty$.

\end{document}